\documentclass[a4paper,11pt]{article}
\usepackage{jheppub} % for details on the use of the package, please see the JINST-author-manual
\usepackage{lineno}
\pdfoutput=1 
\usepackage{subcaption}
\usepackage{graphicx,color}
\usepackage{enumitem}
\usepackage{braket}
\usepackage{slashed}
\usepackage[utf8]{inputenc}
\usepackage{hyperref}
\usepackage{float}
\usepackage{caption}
\usepackage{makecell}
\usepackage{epsfig}
\usepackage{amsmath,amssymb}
\usepackage{hyperref}
\usepackage{diagbox}
\usepackage{lscape}
\usepackage{bbold}
\usepackage{algorithm}
\usepackage{algpseudocode}
\usepackage{amsmath}
\usepackage{amsthm}

\usepackage{mathrsfs}
\usepackage{booktabs}
\usepackage{multirow}

\hypersetup{
	colorlinks=true,
	citecolor=black,
	filecolor=black,
	linkcolor=blue,
	urlcolor=blue
}

\usepackage{dsfont}

\title{\boldmath Rediscovering the Standard Model with AI}

\author[a,b]{Aya Abdelhaq,}
\author[b]{Pellegrino Piantadosi,}
\author[b,c]{Fernando Quevedo}

\affiliation[a]{\footnotesize New York University, 726 Broadway, New York, NY 10003, USA}
\affiliation[b]{New York University Abu Dhabi, PO Box 129188, Saadiyat Island, Abu Dhabi, UAE }
\affiliation[c]{DAMTP, University of Cambridge, Wilberforce Road CB2 0WA, Cambridge, UK}

\emailAdd{aya.abdelhaq@nyu.edu, pellegrino.piantadosi@nyu.edu, fq2054@nyu.edu, fq201@cam.ac.uk}

\abstract{
We investigate whether artificial intelligence can autonomously recover known structures of the Standard Model of particle physics using only experimental data and without theoretical inputs. By applying unsupervised machine learning techniques — including data dimensionality reduction and clustering algorithms — to intrinsic particle properties  and decay modes, we uncover key organizational features of particle physics, such as the relative strength of different interactions and the difference between baryons and mesons. We also identify conserved quantities such as baryon number, strangeness and charm as well as the structure of isospin and the Eightfold Way multiplets. Our analysis then reveals that clustering can separate particles by interaction, flavor symmetries as well as quantum numbers. Additionally, we observe patterns consistent with Regge trajectories in baryon excitations. Our results demonstrate that machine learning can reproduce key aspects of the Standard Model directly from data, suggesting a promising path toward data-driven discovery in fundamental physics.
}

\begin{document}
\maketitle
\flushbottom

\section{Introduction}
\label{sec:intro}

Artificial intelligence (AI) is rapidly establishing itself as a powerful tool across a wide range of scientific disciplines, thanks to its ability to deliver insights and accelerate research. In recent years, machine learning (ML) in particular has found increasing applications in various areas of physics — from experimental high-energy physics to more foundational domains (see for instance \cite{Carleo:2019ptp, cranmer2023interpretablemachinelearningscience, Udrescu:2019mnk,Tsoi:2024pbn, Ruehle:2020jrk, Halverson:2024hax}). This growing relevance raises a natural and important question: can AI contribute  to the formulation of fundamental physical theories?

Before addressing such an ambitious goal, it is essential to test whether AI systems are capable of independently identifying known physical structures from empirical data alone, without any built-in theoretical assumptions. In other scientific fields, encouraging progress has already been made in this direction. In chemistry, for example, unsupervised machine learning techniques have been successfully applied to rediscover the periodic table of elements from data on atomic environments and chemical compounds \cite{Zhou_2018}.

This paper explores whether a similar strategy can be applied to the domain of particle physics. Specifically, we investigate whether unsupervised machine learning techniques can recover the internal structure and symmetries of the Standard Model (SM) — currently the most complete theoretical framework describing the known elementary particles and three of the four fundamental forces — directly from experimental data\footnote{See \cite{Gal:2020dyc,Harvey:2021oue,Butter:2021rvz, Abel:2021ddu,Bendavid:2025urn,Chekanov:2025wzw} for different AI approaches to particle physics phenomenology. In particular, in \cite{Gal:2020dyc} a study  is  performed obtaining properties of baryons from that of mesons and, unlike the present paper, assuming they are composed by quarks. }. Even though our work aims to identify the global symmetries of the Standard Model by just determining conserved charges like baryon number or isospin or flavor $SU(3)$ multiplets, we do not use nor attempt to derive these symmetries. In some sense our work is orthogonal to recent approaches to describe symmetries from machine learning \cite{Krippendorf:2020gny,syvaeri2021improvingsimulationssymmetrycontrol,Wetzel:2020jan}. 

The goal of this work is to assess whether the key organizational features of the Standard Model can emerge naturally from data when analyzed by AI, without the need for human-defined theoretical inputs. To this end, we divide the paper into two parts. In Section \ref{sec:ML_methods}, we introduce the unsupervised learning techniques most relevant for this study, including principal component analysis, t-distributed stochastic neighbor embedding, and clustering algorithms. In Section \ref{sec:applications}, we apply these tools to real particle physics data — focusing first on decay modes and then separately on intrinsic particle properties — to examine whether meaningful structures resembling those of the Standard Model can be recovered.

\section{Data Analysis Techniques}\label{sec:ML_methods}

\subsection{Data dimensionality reduction}

In many real-world datasets, especially in particle physics, the number of variables used to characterize each data point can be large, making it difficult to visualize, interpret, or extract meaningful patterns. Dimensionality reduction refers to a set of techniques that transform high-dimensional data into a lower-dimensional representation while preserving as much relevant structure as possible. This process facilitates both data visualization and downstream analysis by highlighting key relationships and reducing noise or redundancy.

In this section, we focus on two complementary approaches to dimensionality reduction. First, we examine principal component analysis, a linear technique that projects the data onto orthogonal directions of maximal variance. Then, we turn to t-distributed stochastic neighbor embedding, a nonlinear method designed to preserve local neighborhood structures in a reduced space. For more details on these methods, refer to \cite{shlens2014tutorialprincipalcomponentanalysis, JMLR:v9:vandermaaten08a, cai2022theoreticalfoundationstsnevisualizing, arora2018analysistsnealgorithmdata, yin2024rapidreviewclusteringalgorithms}.

\subsubsection*{Principal Component Analysis}\label{sec:PCA}

Principal Component Analysis (PCA) is a statistical method for dimensionality reduction based on linear transformation. Given a dataset $X \in \mathbb{R}^{n,m}$ with $n$ samples and $m$ features, PCA transforms the original correlated variables into a new set of uncorrelated variables, known as principal components. These components are constructed as linear combinations of the original features in such a way that they correspond to the eigenvectors of the covariance matrix of the data. Let $\bar{X}$ be the centered scaled dataset\footnote{Here we refer to the standard scaled dataset, i.e., $\bar{X}_{ij} = \frac{X_{ij} - \mu_j}{\sigma_j}$, where $\mu_j$ and $\sigma_j$ are the mean and standard deviation of feature $j$.}, the covariance matrix is then given by
\begin{equation}
    C=\frac{1}{n-1}\bar{X}^T\bar{X} \ .
\end{equation}
The covariance eigenvectors $v_i$ are obtained by solving the eigenvalue problem
\begin{equation}\label{covariance}
    C v_i = \lambda_i v_i     
\end{equation}
where the eigenvalues $\lambda_i$ indicate the amount of variance captured by each corresponding component.

The principal components are typically ordered by decreasing eigenvalue, so that the first principal component captures the largest amount of variance in the dataset and each subsequent component capturing the maximum remaining variance under the constraint of orthogonality to the other ones. The dimensionality reduction is then achieved by retaining only the first $k$ components, which are sufficient to preserve most of the original variance, and by projecting the data onto the corresponding lower-dimensional subspace: 

\begin{equation}
    X'_k = \bar{X} V_k   
\end{equation}
where $V_k$ is the $m\times k$  matrix constructed with the first $k$ eigenvectors.
In this way, PCA reduces the complexity of the dataset while preserving its most significant variance structure. This facilitates the identification of dominant patterns and correlations within the data, allowing for more effective analysis and interpretation.

It is worth noting that PCA is closely related to the singular value decomposition (SVD) procedure used in \cite{Zhou_2018}. In the case of SVD, the centered data matrix is factorized as
\begin{equation}
    \bar{X} = U \ \Sigma \ V   
\end{equation}
where $U$ and $V$ are, respectively, $n\times n$ and $m \times m$ orthogonal matrices, and $\Sigma = \text{diag}(\sigma_i)$ is a $n \times m$ rectangular diagonal matrix whose entries $\sigma_i$ are the singular values of $\bar{X}$.
Since $\bar{X}^T\bar{X}=V^T\Sigma^TU^TU\Sigma V=V^T\Sigma^T\Sigma V$, the covariance matrix in (\ref{covariance}) is written as
\begin{equation}
    C \equiv\frac{1}{n-1} \ V^T \Sigma^2 \ V \ . 
\end{equation}
The covariance eigenvalues are related to the singular values by $\lambda_i = \frac{\sigma_i^2}{n-1}$, and the dimensionality reduced dataset is
\begin{equation}
    X_k'=U_k \ \Sigma_k
\end{equation}
where $U_k$ and $\Sigma_k$ are obtained considering the first $k$ columns.

\subsubsection*{t-Distributed Stochastic Neighbor Embedding}\label{sec:tsne}

t-Distributed Stochastic Neighbor Embedding (t-SNE) is a nonlinear manifold learning algorithm commonly used for the visualization of high-dimensional datasets. It is specifically designed to capture and preserve local structure in the data, meaning it focuses on maintaining the relative similarity of nearby data points. In contrast to linear dimensionality reduction techniques, which preserve global variance across all features, t-SNE emphasizes the accurate representation of local neighborhoods.

The algorithm constructs a probability distribution over pairs of high-dimensional objects $x_i, x_j \in X$, such that similar objects (close neighborhoods) have a higher probability of being selected. This is typically done by modeling pairwise similarities as Gaussian-distributed conditional probabilities centered at each data point 
\begin{equation}
    p_{j|i}=\frac{\exp{(-|| x_i-x_j||^2/2\sigma_i^2)}}{\sum_{k\neq i} \exp{(-||x_i-x_k||^2/2\sigma_i^2)}} \ .
\end{equation}
Given the conditional probability $p_{j|i}$, the corresponding symmetric joint distribution is obtained as $p_{ij}=\frac{p_{j|i}+p_{i|j}}{2n}$, with $n$ number of data points.

In the low-dimensional embedding, a corresponding joint distribution $q_{ij}$ is defined using a Student t-distribution with one degree of freedom\footnote{The use of a Student t-distribution in the low-dimensional space addresses the tendency of dimensionality reduction to compress large pairwise distances. Unlike a Gaussian, the t-distribution has heavier tails, which allow dissimilar points—those far apart in the high-dimensional space—to remain well-separated in the low-dimensional embedding. },
\begin{equation}\label{tStudent}
    q_{ij}= \frac{(1+||y_i-y_j||^2)^{-1}}{\sum_{k\neq l}(1+||y_k-y_l||^2)^{-1}} \ .
\end{equation}
The goal is to make $q_{ij}$ as similar as possible to $p_{ij}$ in order to preserve the information on the structure of the dataset. This is achieved by minimizing the Kullback-Leibler (KL) divergence between the two distributions:
\begin{equation}
    D_{KL}(p|q) \equiv \sum_{i \neq j}p_{ij} \log{\left( \frac{p_{ij}}{q_{ij}} \right)} \ .
\end{equation}
The divergence acts as a loss function, and the low-dimensional coordinate $y_{i}$ in (\ref{tStudent}) are iteratively updated via a gradient descent method applied on $D_{KL}(p|q)$.
Data points that are close together in high-dimensional space remain close in the embedding, while dissimilar points are modeled to be farther apart.

\subsection{Clustering methods}\label{sec:clustering}
Clustering is an unsupervised machine learning approach aimed at partitioning a dataset into distinct groups (clusters) such that data points within the same group are more similar to each other than to those in different groups. Unlike dimensionality reduction techniques, which project data onto a new coordinate system, clustering algorithms operate directly in the feature space, attempting to uncover the intrinsic grouping structure of the data without prior labeling.

\subsubsection*{$K$-means clustering}\label{sec:kmeans}

$K$-means is one of the most widely used partition-based clustering algorithms. It operates by defining $K$ centroids (cluster centers) and assigning each data point to the nearest centroid, thereby forming $K$ clusters. The algorithm seeks to minimize the within-cluster sum of squares (WCSS):
\begin{equation}\label{eq:cluster_dist}
    D_{WCSS}\equiv\sum_{i=1}^K\sum_{x\in C_i}||x-\mu_i||^2
\end{equation}
where $C_i$ is the set of points in the i-th cluster and $\mu_i$ is the centroid of the cluster. The algorithm follows an iterative refinement approach:
\begin{itemize}
    \item Choose $K$ initial centroids.
    \item Assign each data point to the nearest centroid based on Euclidean distance.
    \item Recompute the centroids as the mean of the points assigned to each cluster.
    \item Repeat tthe previous two steps until convergence.
\end{itemize}
$K$-means assumes spherical, equally sized clusters and may perform poorly on data with non-convex shapes or varying density. It is sensitive to initialization and requires the number of clusters $K$ to be specified in advance.

To quantitatively assess the quality of clustering, one can introduce specific evaluation metrics. In particular, for a given point $i$ one can evaluate the balance between the mean intra-cluster distance (i.e., the average distance to points in the same cluster), $D_{ICi}\equiv\frac{1}{|C_i|-1}\sum_{j\in C_i, j \neq i}D_{ij}$, and the smallest mean distance to the nearest neighboring cluster, $D_{\text{SNC}i} \equiv \min_{C \ne C_i} \left( \frac{1}{|C|} \sum_{j \in C} D_{ij} \right)$, by means of the Silhouette score defined as
\begin{equation}
    S_s = \frac{1}{n}\sum_i S_{si} \equiv \frac{1}{n}\sum_i \frac{D_{SNCi} - D_{ICi}}{\text{max}(D_{ICi}, D_{SNCi})} \ .
\end{equation}
The Silhouette score ranges from $-1$ to $1$: scores $S_s \approx 1$ indicate that the samples are well clustered in terms of cluster separation; values of $S_s \approx 0$ indicate overlapping clusters; negative scores suggest that the points are not properly clustered.

\subsubsection*{Hierarchical clustering}\label{sec:hierarchical_clustering}

Hierarchical clustering is a more flexible technique that produces a nested hierarchy of clusters organized in a tree-like structure (dendrogram). It does not require the number of clusters to be specified a priori, and clusters can be obtained by cutting the dendrogram at a chosen level of granularity.
In the next section, we will make use of the agglomerative approach which starts with each data point as its own cluster and recursively merges the closest pairs of clusters. The closeness between clusters will be defined with the Ward method which consists into minimizing the total within-cluster variance defined in eq. (\ref{eq:cluster_dist}). The algorithm consists in the following steps:
\begin{itemize}
    \item Start with $n$ clusters defined by each point $x_i \in X$.
    \item Compute the within-cluster variance.
    \item Merge the pair of clusters such that the increase in the WCSS is minimized.
    \item Update the within-cluster variance.
    \item Repeat the previous two steps until only a single cluster remains.
\end{itemize}
The iteration results can be visualized via the corresponding dendrogram, where the vertical height at which clusters are merged reflects the dissimilarity between them. By selecting a cut-off height, one can extract a flat clustering with a desired number of clusters or based on distance thresholds. With respect to $K$-means, hierarchical clustering has the advantages of being more informative and interpretable, as it captures multiscale structures and does not require a fixed number of clusters.

\section{Applications in Particle Physics}\label{sec:applications}

In this section, we apply PCA and t-SNE as dimensionality reduction techniques to study patterns in both particle decay channels and intrinsic physical properties. Each particle can be described by a combination of these features, or by the set of possible decay products, forming high-dimensional representations. By reducing these spaces to two or three dimensions, and by applying clustering methods, we aim to uncover underlying structures and groupings that may reflect fundamental particle families, symmetries, and interaction mechanisms. The data used for this analysis are taken from the Particle Data Group (PDG) database \cite{ParticleDataGroup:2024cfk}.

\subsection{Interactions from decay rates}\label{sec:lifetime}

\begin{figure}[h!]
    \centering

    \includegraphics[width=0.83\textwidth]{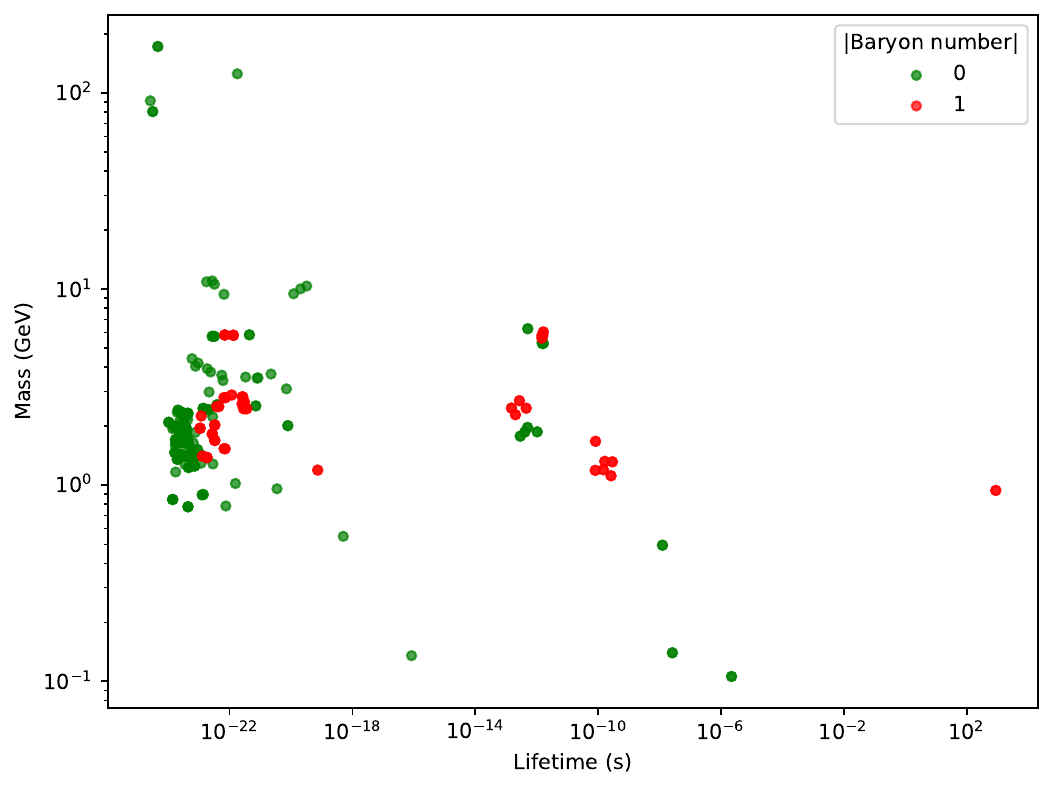}

\caption{Log-plot of mass versus lifetime for known Standard Model particles. Each point represents a particle, with color indicating its baryon number.}
    \label{fig:mass-lifetime}
\end{figure}

Before delving into machine learning results, it is useful to reflect on how much physical insight can be obtained simply by examining basic observables. A clear example is the distribution of Standard Model particles in the mass–lifetime plane \cite{Alimena:2019zri} shown in figure \ref{fig:mass-lifetime}. 
The most immediate observation from this plot is the presence of distinct clusters. One group, found at relatively short lifetimes (order $10^{-22} \ \text{s}$), includes particles that decay predominantly via the strong interaction. These are mainly hadrons whose decay is driven by the large coupling strength of the strong force. The strong interaction leads to extremely rapid decay processes, consistent with the short lifetimes observed. Even though quarks and gluons are never seen in isolation due to confinement, their presence is indirectly encoded in the decay properties of hadrons.
In contrast, a second, more diffuse cluster appears at longer lifetimes, ranging from $10^{-12}$ to $10^{-6}$ seconds. These particles decay primarily through the weak interaction, which is several orders of magnitude weaker than the strong force. As a result, their decay processes are rarer and occur on longer timescales. This group includes charged leptons, neutrinos, and weakly decaying hadrons. The pattern is particularly striking because it emerges without any explicit labeling by interaction type; it arises naturally from physical decay data.

This simple plot therefore serves as a compelling demonstration of how particle properties reflect underlying interaction dynamics. Particles subject to strong decays cluster separately from those governed by weak decays, solely on the basis of lifetime. 
\begin{figure}[h!]
    \centering

    \includegraphics[width=0.8\textwidth]{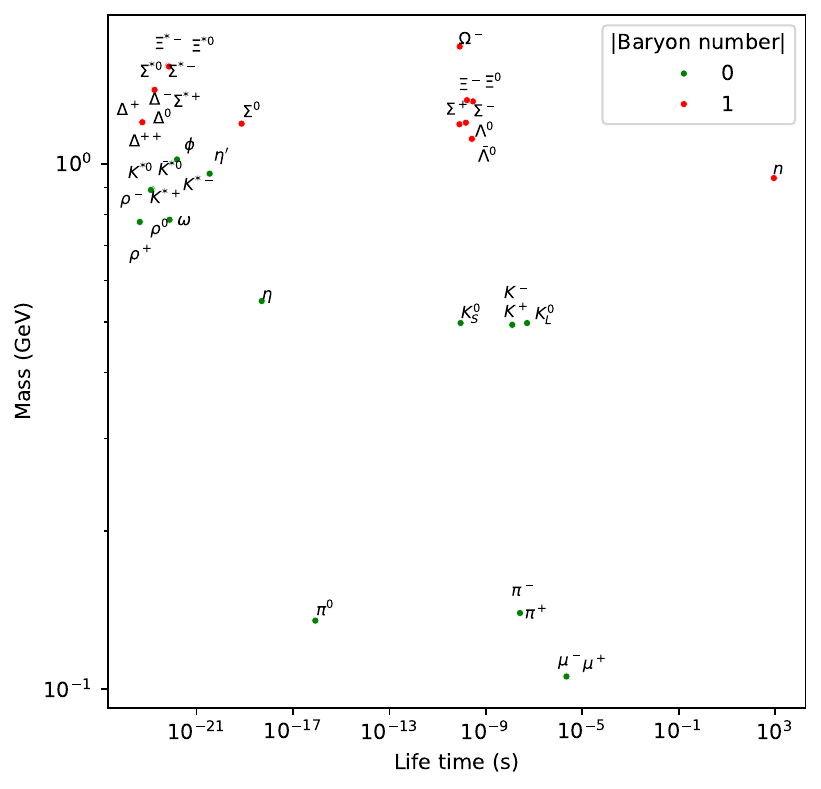}

\caption{Mass versus lifetime for light mesons and baryons known at the time of the Eightfold Way. Colors indicate baryon number. A clear separation in mass is visible: all baryons have masses equal to or greater than that of the proton and neutron, while mesons lie below this threshold.}

    \label{fig:mass-lifetime_restricted}
\end{figure}

Figure \ref{fig:mass-lifetime_restricted} shows the same plot limited to the particles known before $1960$. It is worth noting that $\pi_0$, $\eta$ mesons and $\Sigma^0$ appear between the two main clusters, reflecting their intermediate lifetimes. This is consistent with the fact that both particles primarily decay via electromagnetic interactions. We also observe that baryons and mesons can be distinguished based on their mass. Specifically, all baryons have a mass equal to or greater than that of the proton, while mesons have masses that range between the electron and proton masses. However, this classification criterion no longer holds once the particles discovered later are included. It is therefore interesting to explore whether baryons and mesons can still be classified using alternative methods based solely on experimental observations. This possibility will be examined in the following section.

\subsection{Baryon-meson classification}

The clear clustering discussed in the previous section motivates the use of unsupervised machine learning techniques to uncover more subtle patterns when higher-dimensional information is taken into account. In this and next sections, we build on this idea, showing how such techniques can recover and even extend our understanding of particle classification based on physical properties alone.

In the context of particle physics, each particle decays into various products, such as leptons, neutrinos, or hadrons. These decay products form a set of features that can be used to describe the particle’s decay mode. More precisely, the input to the machine can be structured as a dataframe with particles as rows and particles as columns. For each row particle, we examine the decay products across all of its decay modes. The entry at a given row–column position is set to $1$ if the column particle appears as a decay product in at least one of the row particle’s decay modes, and $0$ otherwise. The high-dimensional space generated by these features can be difficult to interpret directly, but PCA described in Section \ref{sec:PCA} allows us to reduce it to a few dimensions, making it easier to identify clusters and patterns within the data. In this specific case, due to the binary nature of the dataset, we apply a row vector normalization\footnote{Instead of the standard scaling considered in Section \ref{sec:PCA}, here we adopt a normalisation of the form $\bar{X}_{ij} = \frac{X_{ij}-\mu_j}{\sqrt{\sum_j (X_{ij}-\mu_j)^2}}$.} as pre-processing, in order to avoid biasing the PCA toward particles with a larger number of decay modes.

\begin{figure}[h!]
    \centering

    \includegraphics[width=0.8\textwidth]{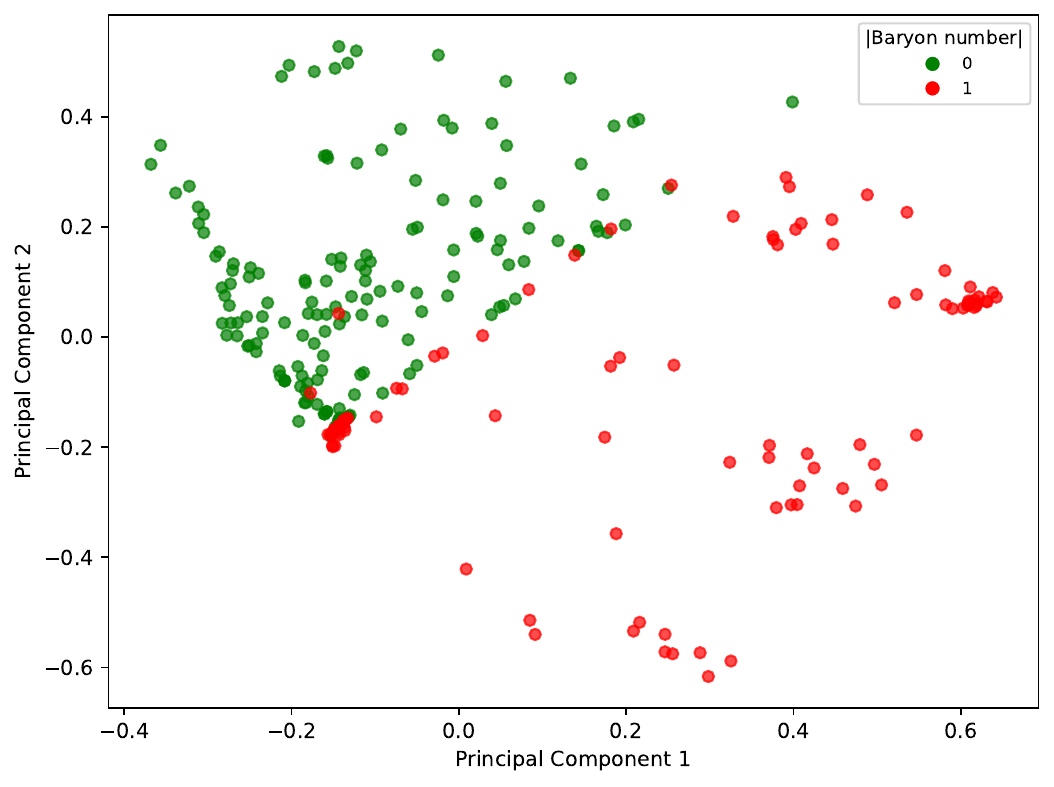}

\caption{Projection of particle decay modes onto the first two principal components from PCA applied to particles decay modes. Points are colored by baryon number, clearly separating baryons from mesons and illustrating distinct decay pattern groupings based on baryon classification.}
    \label{fig:pca}
\end{figure}

The plot shown in Figure \ref{fig:pca} illustrates the first two principal components of the decay product channels, where each data point corresponds to a particle, and the position of the particle on the plot reflects its decay channel features projected onto the first two principal components. These two components capture the majority of the variance in the dataset, making them the most significant dimensions for understanding the relationships between particles and their decay behaviors.

Although two distinct clusters are not observed, it is evident that the particles are separated by a straight boundary, with baryons concentrated below the line and mesons above it. The two baryonic outliers appearing in the upper (mesonic) region correspond to the proton and neutron, for which decay-mode-based analysis is clearly ineffective, due to their stability and lack of accessible decay channels under normal conditions.

A complementary approach to studying particle decay properties involves analyzing the entire decay chain, including not only the products of the primary decay channel but also those resulting from secondary decays. The goal of this method is to provide the machine with information about the internal structure of each particle—specifically, how it breaks down into more stable particles. In practice, this means constructing again an input matrix where each row corresponds to a particle and each column represents a possible decay product—that is, one column for each known particle. For a given particle (row), the matrix entries are set to $1$ or $0$ depending on whether the corresponding particle (column) appears among the primary or secondary decay products of that particle.

\begin{figure}[h!]
    \centering

    \includegraphics[width=0.75\textwidth]{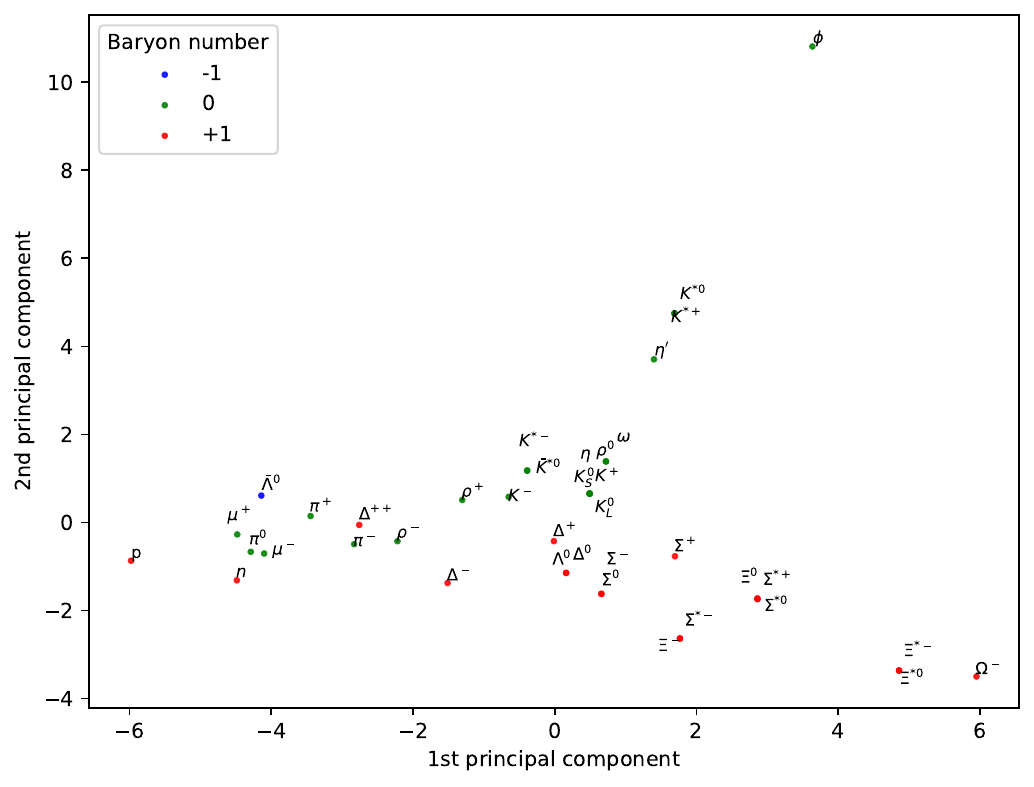}

\caption{Second vs first principal component from PCA applied to the full decay chain products of light baryons and mesons. The separation between baryons and mesons reflects differences in the structure and composition of their decay cascades.}
    \label{fig:meson-baryon-chain-res}
\end{figure}

\begin{figure}[h!]
    \centering

    \includegraphics[width=0.8\textwidth]{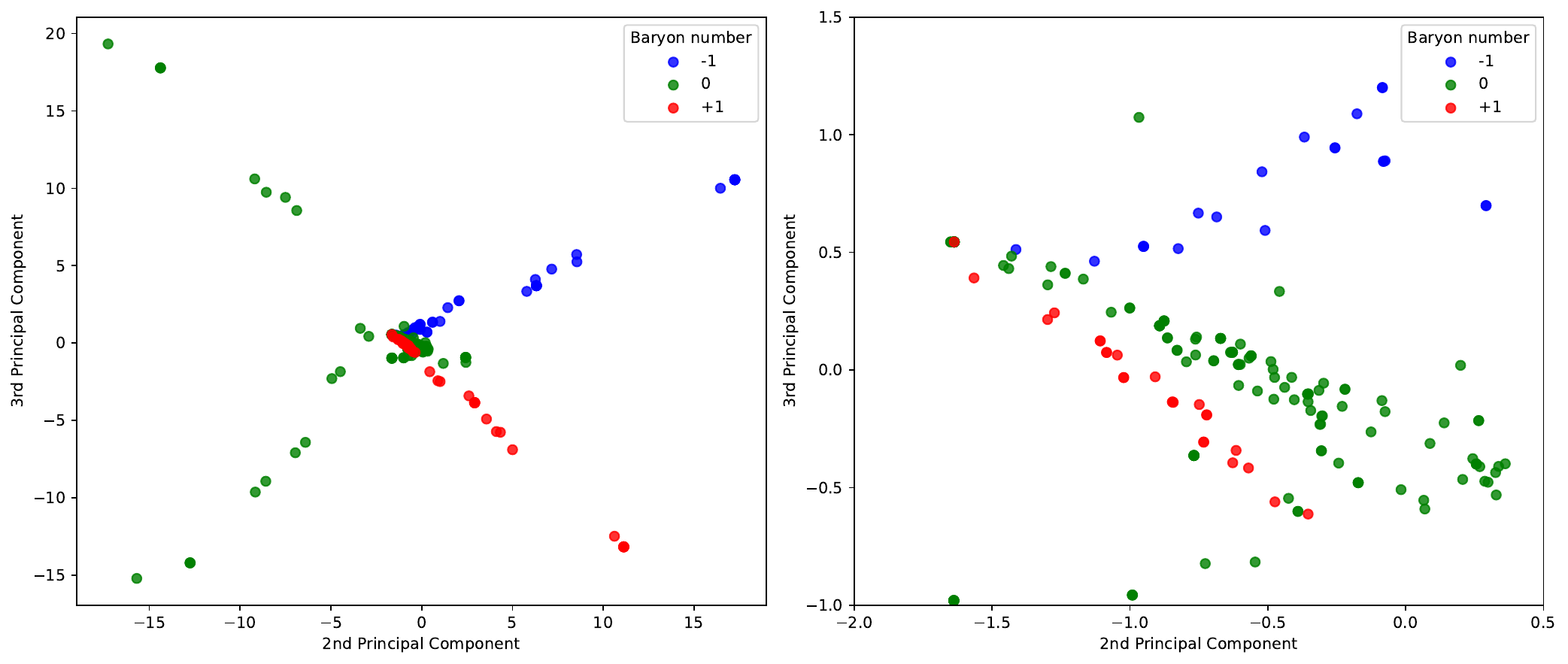}

\caption{\textit{Left}: PCA projection onto the second and third components using decay product data from all known PDG-listed particles. The plot shows clear linear alignments separating baryons, mesons, and antibaryons, reflecting underlying differences in their decay product patterns. \textit{Right}: Zoom-in on the dense accumulation region.}
    \label{fig:meson-baryon-chain}
\end{figure}

The results are shown in Figures \ref{fig:meson-baryon-chain-res}, \ref{fig:meson-baryon-chain}. Once again, and even more distinctly than before, the particles are seen to split based on their baryonic or mesonic nature. Figure \ref{fig:meson-baryon-chain-res} shows the restricted dataset discussed in the previous section, while Figure \ref{fig:meson-baryon-chain} includes all known particles. In this extended case, the split captures not only the baryon/meson classification but also the distinction between baryons ($B=+1$) and antibaryons ($B=-1$). This successful separation arises from the fact that, at the end of their decay chains, baryons invariably produce neutrons and/or protons, while mesons typically decay into leptons. The method described here is clearly unsupervised, which aligns well with the goals of this study.

\subsection{$SU(N)$ flavour multiplets}

To reproduce more complex structures related to the symmetries of the Standard Model, it is necessary to include additional information beyond just mass and lifetime. A natural generalization of the case considered in Section \ref{sec:lifetime} is to incorporate spin information into the analysis. Spin, moreover, is an experimentally observable quantity, aligning with our approach of reconstructing the Standard Model using purely empirical data. Given the limited number of features in this case, and considering that most of the variance would be dominated by the spin, it is more appropriate to explore non-linear relationships among the features using t-SNE, as described in Section \ref{sec:tsne}.

At this point, one can reconstruct the historical path of particle discoveries, beginning by restricting the analysis to the lightest hadrons composed of up and down quarks only\footnote{Note that even though we refer to 'quarks' as a simple way of describing hadrons, it is important to emphasize that we are {\emph not} assuming at any stage that hadrons are composite.}, along with all the leptons. In the absence of any prior knowledge of quark substructure, the experimental motivation for selecting this subset lies in focusing on the lightest hadrons that do not exhibit the \textit{strange} behavior typical of hadrons containing strange quarks — namely, particles that are produced via the strong interaction but decay with unexpectedly long lifetimes. Leptons, on the other hand, can be distinguished by the fact that they decay via the weak interaction (as seen in the long-lifetime cluster in Figure \ref{fig:mass-lifetime}) and are also typically produced in hadronic decays through channels with relatively small branching ratios.
\begin{figure}[h!]
    \centering

    \includegraphics[width=0.6\textwidth]{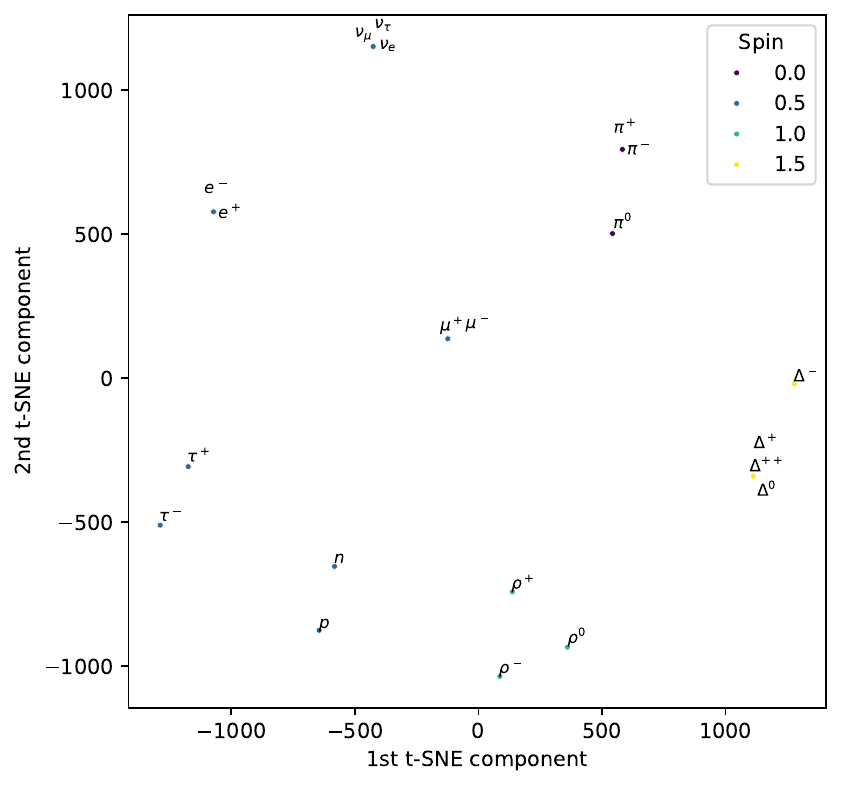}

\caption{t-SNE projection of baryons and mesons based on their physical properties (mass, life-time and spin) restricted to only light hadrons and leptons. Points are colored by spin. The clustering reveals the structure of SU(2) flavor multiplets, with spin helping to distinguish each group.}
    \label{fig:Multiplets_res_res}
\end{figure}

Despite the limited number of particles considered, small clusters can still be distinguished which, in the case of hadrons, correspond precisely to the $SU(2)$ isospin multiplets \footnote{Again, even though we mention $SU(2)$ and $SU(3)$ flavor symmetries it should be clear that the existence of these symmetries is {\emph not} assumed in this work.}. For example, the proton\footnote{It should be noted that for the proton, an arbitrary large lifetime of the order of $10^{50} s$ was assigned.} and neutron are grouped together, as are the $\Delta$ particles, and separately, the $\pi$ and $\rho$ mesons. 

\begin{figure}[h!]
    \centering

    \includegraphics[width=0.8\textwidth]{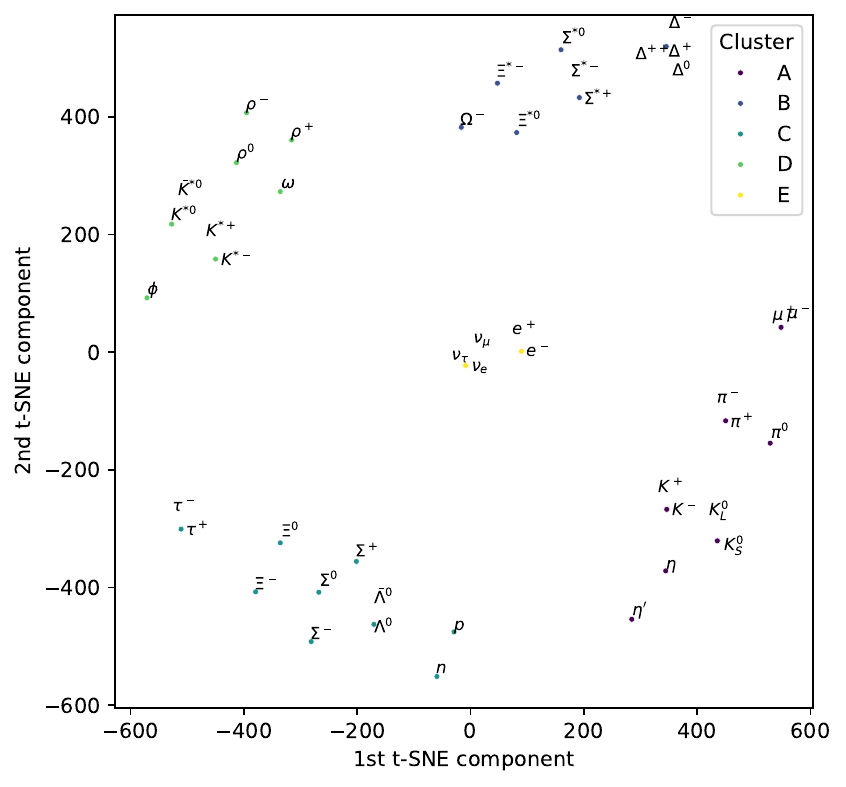}

\caption{t-SNE projection of baryons, mesons and leptons based on their physical properties (mass, life-time and spin). Points are colored according to $K$-means cluster labels. The clustering reveals the structure of SU(3) flavor multiplets.}
    \label{fig:Multiplets}
\end{figure}

If we continue to focus on the particles known before $1960$ and include spin in the analysis, it becomes interesting to apply the dimensionality reduction and clustering techniques introduced in the previous section. Figure \ref{fig:Multiplets} shows the plot of the first two components obtained by applying t-SNE to the dataset that includes again mass, spin, and lifetime as input features (see Table \ref{tab:particle_multiplets}). The colors indicate the different labels obtained by applying $K$-means clustering for $K=5$.
\begin{table}[htbp]
\centering
\resizebox{\textwidth}{!}{
\begin{tabular}{lccccccc}
\toprule
\textbf{Particle} & \textbf{Mass [MeV/$c^2$]} & \textbf{Spin $J$} & \textbf{Charge [$e$]} & \textbf{Baryon Number $B$} & \textbf{Isospin $I$} & \textbf{Lifetime [s]} & \textbf{SU(3) Multiplet} \\
\midrule
$\pi^0$         & 134.98   & 0     & 0   & 0     & 1     & $8.4 \times 10^{-17}$ & \multirow{9}{*}{$0^-$ Meson Nonet}\\
$\pi^+$         & 139.57   & 0     & +1  & 0     & 1     & $2.6 \times 10^{-8}$ &  \\
$\pi^-$         & 139.57   & 0     & -1  & 0     & 1     & $2.6 \times 10^{-8}$ & \\
$K_S^0$         & 497.61   & 0     & 0   & 0     & 1/2   & $8.9 \times 10^{-11}$ & \\
$K_L^0$         & 497.61   & 0     & 0   & 0     & 1/2   & $5.1 \times 10^{-8}$ & \\
$K^+$           & 493.68   & 0     & +1  & 0     & 1/2   & $1.2 \times 10^{-8}$ & \\
$K^-$           & 493.68   & 0     & -1  & 0     & 1/2   & $1.2 \times 10^{-8}$ & \\
$\eta$          & 547.86   & 0     & 0   & 0     & 0   & $5.0 \times 10^{-19}$ & \\
$\eta'$         & 957.78   & 0     & 0   & 0     & 0   & $3.2 \times 10^{-21}$ & \\
\midrule

$p$             & 938.27   & 1/2   & +1  & 1     & 1/2   & Stable               & \multirow{8}{*}{$1/2^+$ Baryon Octet} \\
$n$             & 939.57   & 1/2   & 0   & 1     & 1/2   & $8.2 \times 10^2$    & \\
$\Lambda^0$     & 1115.68  & 1/2   & 0   & 1     & 0     & $2.6 \times 10^{-10}$ & \\
$\Sigma^0$      & 1192.64  & 1/2   & 0   & 1     & 1     & $7.4 \times 10^{-20}$ & \\
$\Sigma^+$      & 1189.37  & 1/2   & +1  & 1     & 1     & $8.0 \times 10^{-11}$ & \\
$\Sigma^-$      & 1197.45  & 1/2   & -1  & 1     & 1     & $1.5 \times 10^{-10}$ & \\
$\Xi^0$         & 1314.86  & 1/2   & 0   & 1     & 1/2   & $2.9 \times 10^{-10}$ & \\
$\Xi^-$         & 1321.71  & 1/2   & -1  & 1     & 1/2   & $1.6 \times 10^{-10}$ & \\
\midrule

$\rho^0$        & 775.26   & 1     & 0   & 0     & 1     & $4.5 \times 10^{-24}$ & \multirow{9}{*}{$1^-$ Meson Nonet}\\
$\rho^+$        & 775.11   & 1     & +1  & 0     & 1     & $4.4 \times 10^{-24}$ & \\
$\rho^-$        & 775.11   & 1     & -1  & 0     & 1     & $4.4 \times 10^{-24}$ & \\
$\omega$        & 782.65   & 1     & 0   & 0     & 0     & $7.6 \times 10^{-23}$ & \\
$K^{*0}$        & 895.81   & 1     & 0   & 0     & 1/2   & $1.4 \times 10^{-23}$ & \\
$\bar{K}^{*0}$  & 895.81   & 1     & 0   & 0     & 1/2   & $1.4 \times 10^{-23}$ & \\
$K^{*+}$        & 891.66   & 1     & +1  & 0     & 1/2   & $1.3 \times 10^{-23}$ & \\
$K^{*-}$        & 891.66   & 1     & -1  & 0     & 1/2   & $1.3 \times 10^{-23}$ & \\
$\phi$          & 1019.46  & 1     & 0   & 0     & 0     & $1.5 \times 10^{-22}$ &  \\
\midrule

$\Delta^{++}$   & 1232     & 3/2   & +2  & 1     & 3/2   & $5.6 \times 10^{-24}$       & \multirow{10}{*}{$3/2^+$ Baryon Decuplet} \\
$\Delta^{+}$    & 1232     & 3/2   & +1  & 1     & 3/2   & $5.6 \times 10^{-24}$       & \\
$\Delta^{0}$    & 1232     & 3/2   & 0   & 1     & 3/2   & $5.6 \times 10^{-24}$       & \\
$\Delta^{-}$    & 1232     & 3/2   & -1  & 1     & 3/2   & $5.6 \times 10^{-24}$       & \\
$\Sigma^{*+}$   & 1382.8   & 3/2   & +1  & 1     & 1     & $4.48 \times 10^{-23}$ & \\
$\Sigma^{*0}$   & 1383.7   & 3/2   & 0   & 1     & 1     & $4.45 \times 10^{-23}$ & \\
$\Sigma^{*-}$   & 1387.2   & 3/2   & -1  & 1     & 1     & $4.35 \times 10^{-23}$ & \\
$\Xi^{*0}$      & 1531.8   & 3/2   & 0   & 1     & 1/2   & $2.3 \times 10^{-23}$ & \\
$\Xi^{*-}$      & 1535.0   & 3/2   & -1  & 1     & 1/2   & $2.1 \times 10^{-23}$ & \\
$\Omega^-$      & 1672.45  & 3/2   & -1  & 1     & 0     & $8.2 \times 10^{-11}$ & \\
\bottomrule
\end{tabular}
}
\caption{Properties of particles grouped by their corresponding $SU(3)$ multiplets.}
\label{tab:particle_multiplets}
\end{table}

\begin{figure}[h!]
    \centering

    \includegraphics[width=0.8\textwidth]{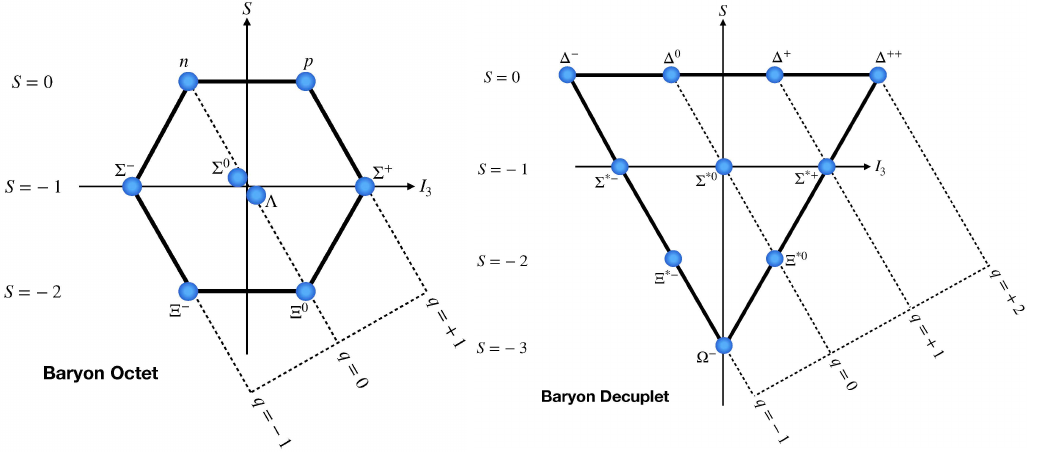}

\caption{\textit{Left}: $1/2^+$ spin-parity baryon $SU(3)$ octet. \textit{Right}: $3/2^+$ spin-parity baryon $SU(3)$ decuplet. \textit{Images adapted from \cite{Quevedo:2024kmy}}.}
    \label{fig:Eightfoldway}
\end{figure}

We observe that the hadronic particles tend to group according to the flavor $SU(3)$ multiplets, partially shown in Figure \ref{fig:Eightfoldway}, which were historically introduced under the name of \textit{Eightfold Way}. In particular, we recover the classification into baryon decuplet and octet representation with spin-parity $\frac{1}{2}^{+},\frac{3}{2}^{+}$ according to the well-known decomposition in terms of the quarks constituents\footnote{Again we use the quark composition for clarity of explanation but we are not assuming any symmetry nor representation, we find the clusters by AI identifying particles with similar porperties.}:
\begin{equation}
    \text{Baryons:} \quad3\otimes3\otimes3=10_{\mathcal{S}}\oplus 8_{\mathcal{M}}\oplus 8_{\mathcal{M}} \oplus 1_{\mathcal{A}}
\end{equation}
and the meson octets (nonets) with spin-parity $0^-, 1^-$, in the decomposition
\begin{equation}
    \text{Mesons:} \quad3\otimes\bar{3}=8 \oplus 1 \ .
\end{equation}
At first glance, one might assume that this clustering is to be expected, as spin alone might be sufficient to distinguish between these multiplets. However, to test the robustness of the method, we also included leptons, which of course do not belong to the aforementioned multiplets. As expected, these particles do not fall into the same spin cluster: the lighter leptons form a separate cluster (electron and neutrinos), whereas the muon and tau lepton fall outside this grouping, evidently due to their larger mass differences.
By computing the Silhouette score, introduced in Section \ref{sec:kmeans}, we find that after dimensionality reduction it reaches $S_s=0.6$, compared to $0.5$ for the original dataset. This indicates a good cluster separation in multiplets that emerges more clearly from the reduction analysis. 

\begin{figure}[h!]
    \centering

    \includegraphics[width=0.9\textwidth]{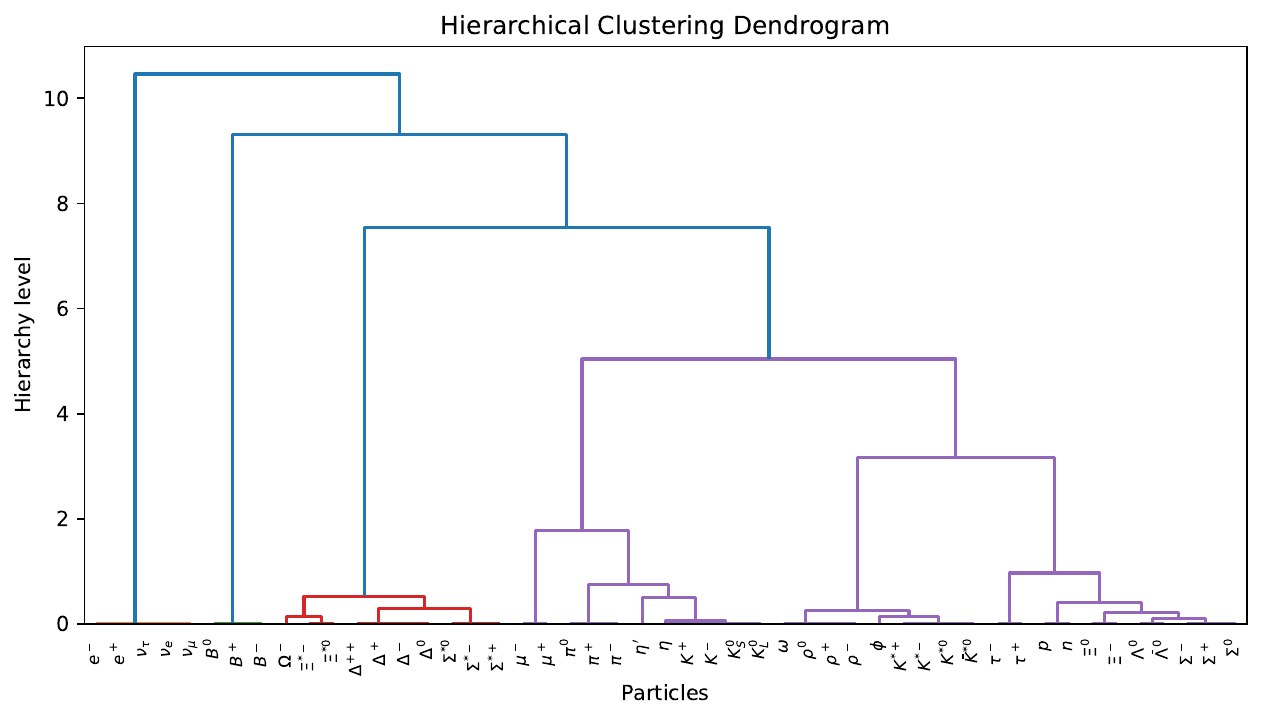}

\caption{Hierarchical clustering dendrogram of particles revealing groupings consistent with the SU(3) flavor structure. The dendrogram’s sublevels correspond to finer subdivisions that reflect the isospin multiplet organization within each SU(3) multiplet.}
    \label{fig:Hierarchy}
\end{figure}

An alternative, fully unsupervised approach that does not require predefining the number of clusters is hierarchical clustering described in Section \ref{sec:hierarchical_clustering}.
For our analysis, we used the agglomerative approach, which is more commonly applied when the goal is to reveal the structure of the data from the bottom-up. The similarity between the data points (particles, in this case) is measured using a distance metric based on their mass, spin and lifetime, and the clustering is performed using the Ward’s linkage method.
The resulting hierarchical clustering dendrogram in Figure \ref{fig:Hierarchy}, visually represents the way the particles are grouped. This clustering not only separates particles into broad categories but also reveals finer substructures within each group. Specifically, it reveals a classification that mirrors the organization into ispospin $SU(2)$ multiplets, highlighting the internal symmetry structure within each $SU(3)$ cluster.

The same type of analysis can be carried out to reproduce the $SU(4)$ symmetry multiplets corresponding to the following decompositions:
\begin{equation}
    \text{Baryons:} \quad4\otimes4\otimes4=20_{\mathcal{S}}\oplus 20_{\mathcal{M}}\oplus 20_{\mathcal{M}} \oplus 4_{\mathcal{A}}
\end{equation}
\begin{equation}
    \text{Mesons:} \quad4\otimes\bar{4}= 15 \oplus 1 \ .
\end{equation}

\begin{figure}[h!]
    \centering

    \includegraphics[width=0.9\textwidth]{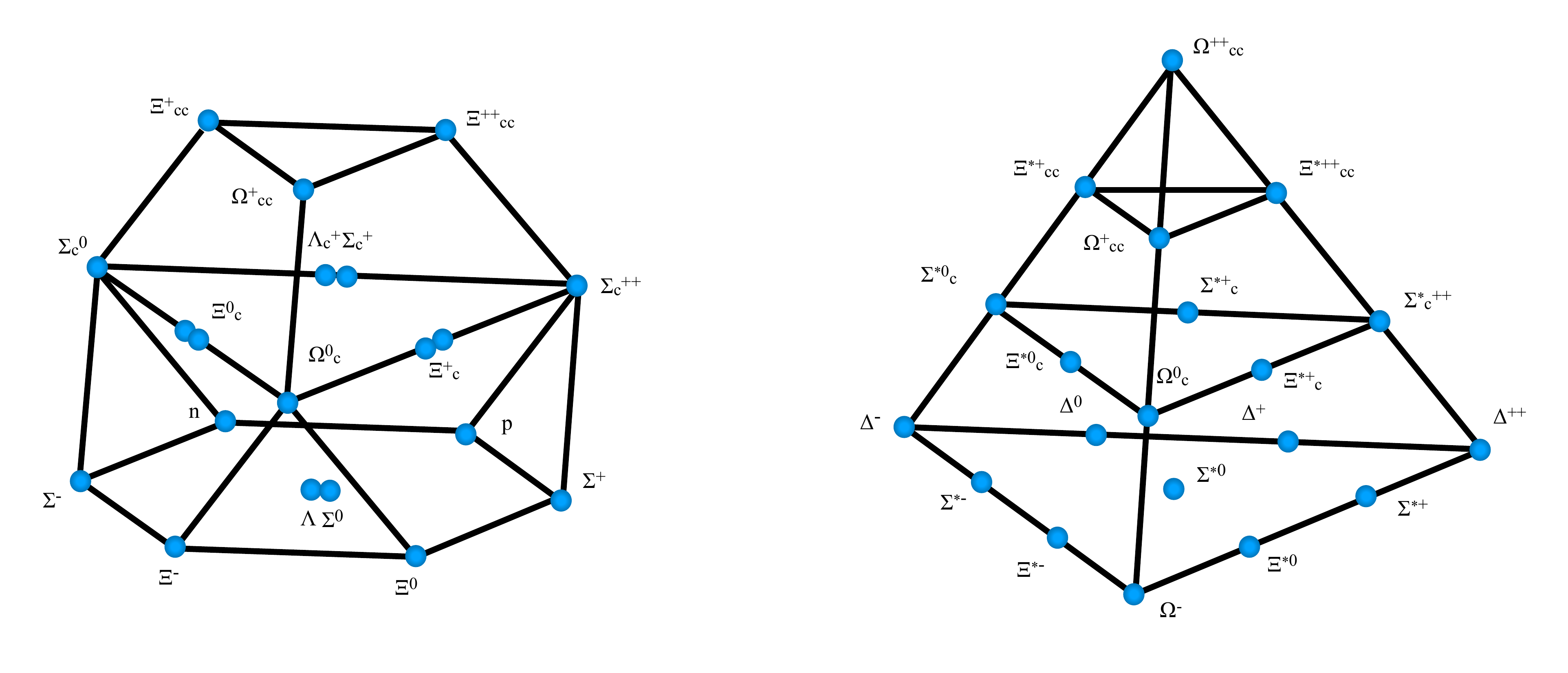}

\caption{\textit{Left}: $1/2^+$ spin-parity baryon $SU(4)$ $20$-plet. \textit{Right}: $3/2^+$ spin-parity baryon $SU(4)$ 20-plet.}
    \label{fig:su4_multiplets}
\end{figure}

\begin{figure}[h!]
    \centering

    \includegraphics[width=0.9\textwidth]{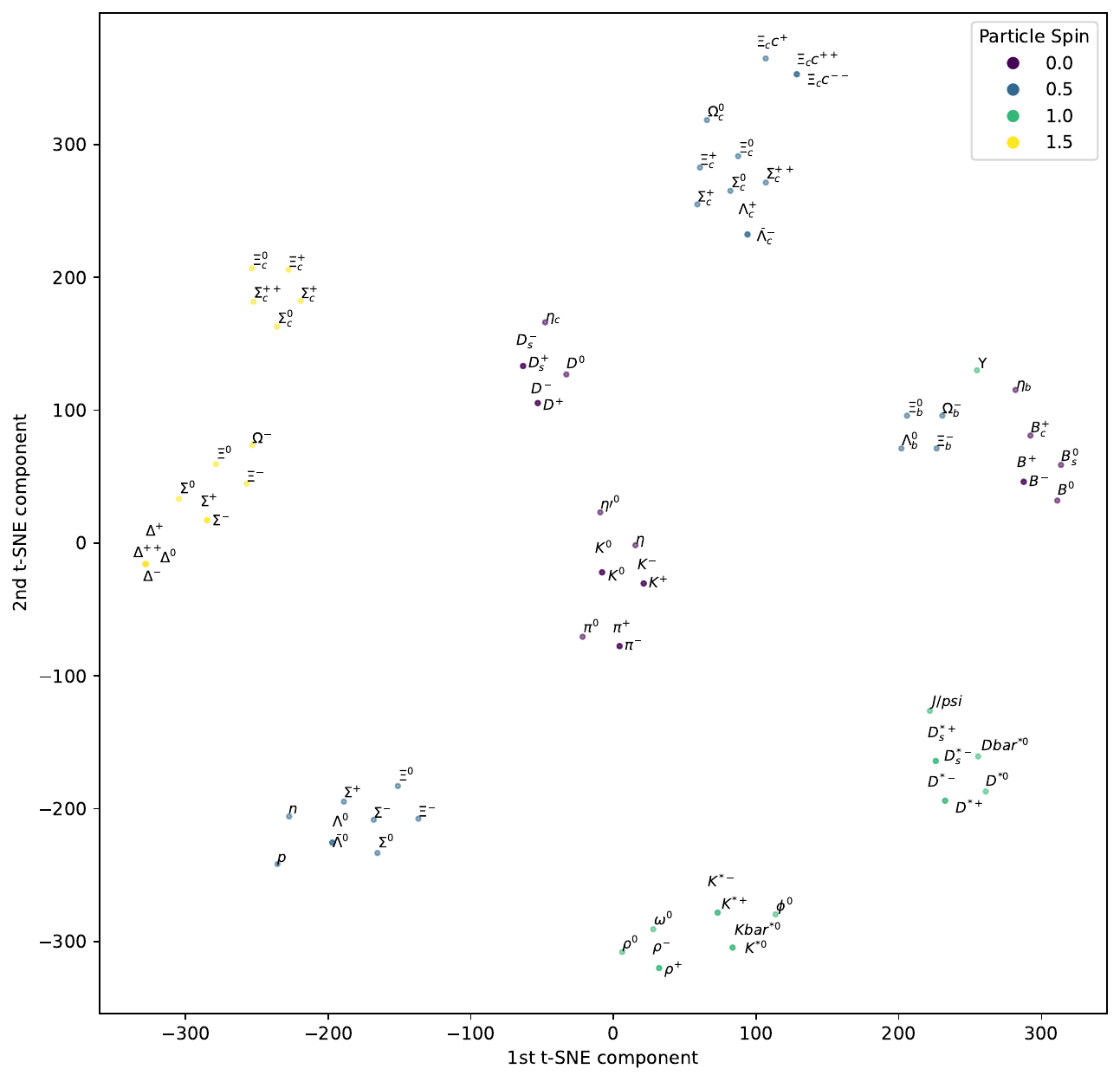}

\caption{t-SNE plot of light and heavy hadrons based on their physical properties. Distinct clusters correspond to $SU(3)$ flavor multiplets for the light particles, while charmed baryons and mesons form groups consistent with $SU(4)$ multiplet structure, highlighting the extension of flavor symmetry to include the charm quantum number.}
    \label{fig:tsne-hierarchy-su4}
\end{figure}

In this case, the results are shown in Figure \ref{fig:tsne-hierarchy-su4}. The goal is to identify groupings that correspond to well-known structures, which are included for reference in Figure \ref{fig:su4_multiplets}.

When the charm quantum number is introduced, the flavor symmetry group is extended from 
$SU(3)$ to $SU(4)$, where each $SU(4)$  multiplet encompasses multiple $SU(3)$ multiplets distinguished by their charm content. In particular, note that in presence of a charm quantum number, one can identify baryon $SU(3)$ subgroups corresponding  to a sextet or an anti-triplet (see \cite{Crede:2013kia} for details). In our case, the most distinct clusters again align with $SU(3)$ multiplets, reflecting well-defined flavor structures. Moreover, we observe that $SU(3)$ multiplets belonging to the same $SU(4)$ representation tend to lie closer to each other in the reduced space. This hierarchical arrangement is also supported by the clustering results, suggesting that $SU(4)$ symmetry is still visible, albeit more weakly. This reduced clarity is consistent with the fact that $SU(4)$ is only an approximate symmetry, broken by the relatively large mass of the charm quark, which lies close to the QCD scale.

\subsection{Regge trajectories}

In the previous section, we examined how to group together particles with the same spin and similar mass. We now turn to the question of how to relate particles that share the same quantum numbers but differ in mass and orbital angular momentum. In other words, our aim is to study the structure of the Standard Model in the context of excited states \footnote{We thank Luis Ib\'a\~nez for encouraging us to explore this avenue.}.
Spin and mass turn out to be not independent quantities. According to the Regge theory for hadron scattering \cite{Collins_2023}, it is possible to establish an approximate linear relation between the total angular momentum $J$ and the squared mass $m^2$ of the particles:
\begin{equation}\label{eq:regge}
    J=\alpha'm^2 + J_0 \ 
\end{equation}
where $\alpha'$ is the so-called \textit{Regge slope}, which is almost the same for all the particles, and the intercept $J_0$ defines a single \textit{Regge trajectory}. In principle, given a trajectory specified by $\alpha'$, $J_0$, it could be populated at discrete positive values of total angular momentum, either integer or half-integer, depending on whether the particles are mesons or baryons.

Regge trajectories thus offer a useful classification scheme for organizing the spectrum of hadronic resonances. Particles lying on the same trajectory are interpreted as orbital excitations of the lightest (or ground-state) particle in the series.

\begin{figure}[h!]
    \centering

    \includegraphics[width=0.9\textwidth]{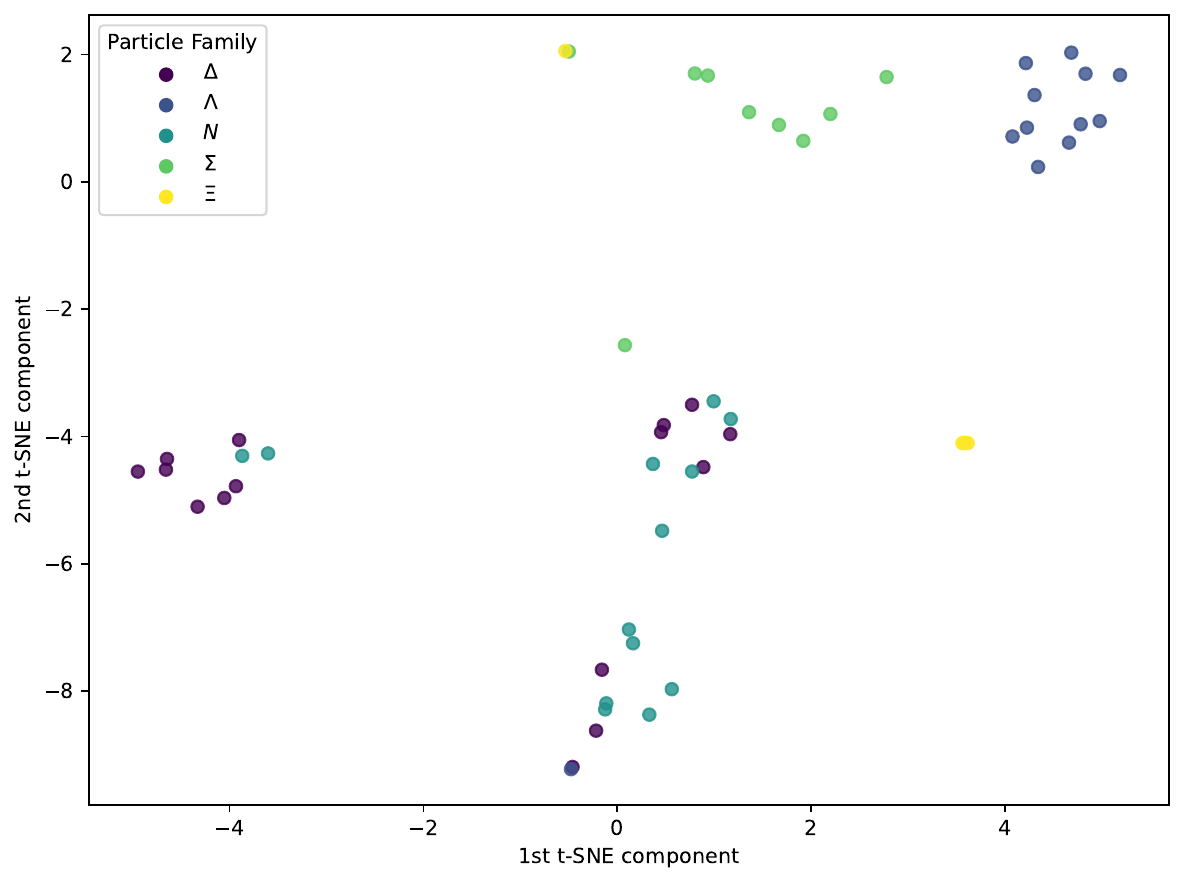}

\caption{ t-SNE dimensionality reduction applied to the decay-mode dataset for excited baryons. The projection reveals well-separated clusters corresponding to the $\Sigma, \Lambda$, and $\Xi$ families, while nucleons and $\Delta$ baryons form two broader, partially overlapping clusters, reflecting similarities in their decay patterns. Notably, a few outlier points lie detached from the main groupings; these correspond to the ground states of the baryons, whose decay behavior is qualitatively distinct from their excited counterparts.}
    \label{fig:excitations}
\end{figure}
We want to explore how Regge trajectories can be at least qualitatively reproduced by the machine using experimental data alone. For simplicity, we restrict the analysis on the classification of orbital excitations within baryonic families for which the trajectory structures are more clearly established. The first step in the procedure involves the unsupervised clustering of baryonic resonances using their decay modes as features. For each particle, decay modes are weighted by the respective branching ratios to emphasize the main contributions. The underlying hypothesis is that particles lying on the same Regge trajectory, as orbital excitation of the same ground state, will share dominant decay patterns.  To ensure robustness of the classification, only particles with well-measured mass, total angular momentum and observed decay are considered.

The dimensionality reduction, shown in Figure \ref{fig:excitations}, yields distinct clusters corresponding to the known baryonic families. In particular, $\Lambda$, $\Sigma$ and $\Xi$ baryons form separate, clearly distinguishable clusters; nucleons $N$ and $\Delta$ resonances tend to group together, reflecting their shared quark content and common dominant decay mode $N + \pi$. Outlier particles are found to correspond to ground states, which naturally lack the decay structure typical of their excited states.

After clustering, each group of particles can be analyzed to construct $J$ vs $m$ plots from which we observe that:
\begin{itemize}
    \item Particles align along approximately parabolic curves, consistent with the expectations for Regge trajectories.
    
    \item Each curve begins with the ground state and contains its higher orbital excitations.

    \item The alignment allows for \textit{a posteriori} identification of the outlier ground state in the corresponding excitation cluster.

    \item The $J$ vs $m$ representation provides the separation between nucleons and $\Delta$ baryons which were overlapping in the dimensionality reduced decay mode space.
    
\end{itemize}

An important observation is that the unsupervised classification correlates more strongly with the isospin than with Regge behavior. For a given isospin, we can have multiple Regge trajectories \cite{Klempt:2012fy}:
\begin{itemize}
    \item A leading Regge trajectory, composed of the lowest-mass states at each $J$.

    \item One or more daughter trajectories, arising from radial excitations in the particle wave function or spin-orbit splittings.
    
\end{itemize}
For simplicity, we restrict the analysis to the leading trajectories only, selecting the lightest particle for each value of $J$. Each trajectory is then fitted using a quadratic relation.
\begin{figure}[h!]
    \centering

    \includegraphics[width=0.7\textwidth]{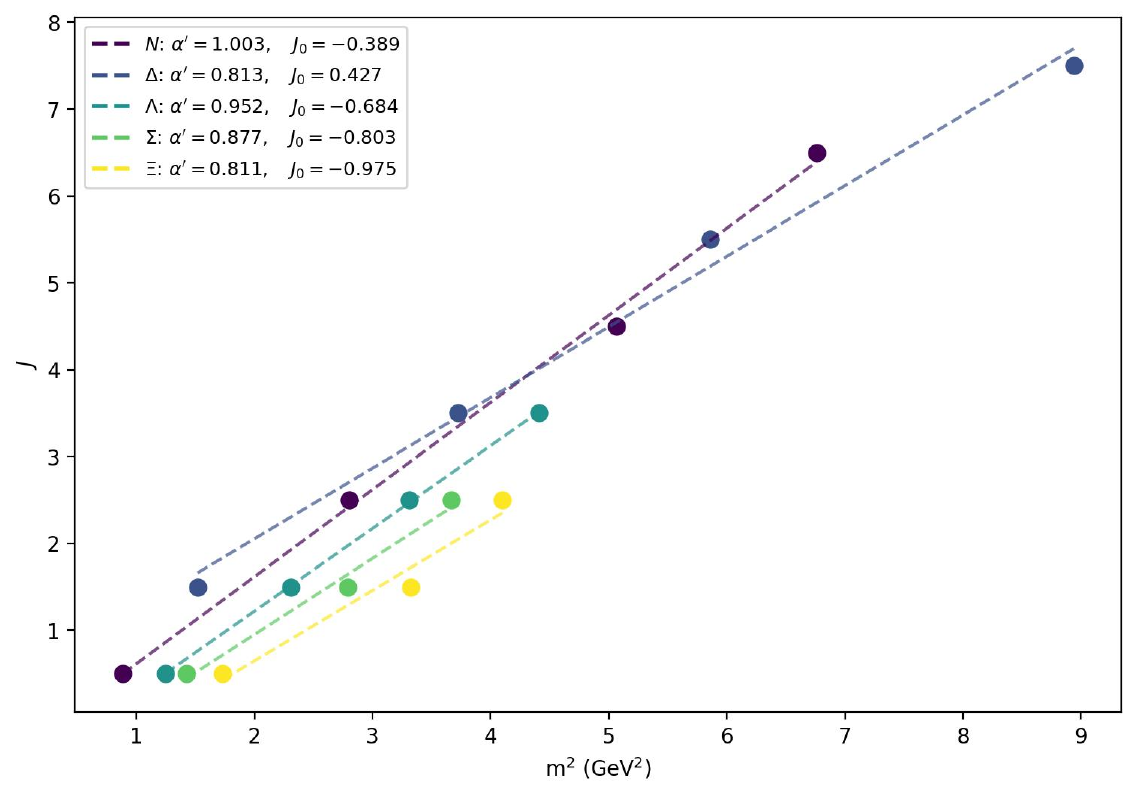}

\caption{Regge trajectories for baryons plotted as total angular momentum $J$ vs $m^2$. The data points corresponding to different baryon families align along approximately linear trajectories, each with a similar slope $\alpha' \approx 0.9 \ \text{GeV}^{-2}$, consistent with the expected behavior from Regge theory.}
    \label{fig:regge}
\end{figure}
The fits indicate that the linear term is negligible, supporting a predominantly quadratic dependence on mass. Moreover, the coefficient of the quadratic term, the slope parameter in eq. (\ref{eq:regge}), appears to be approximately universal across different baryon families, as expected. This leads to a clear linear alignment of the states in the $J-m^2$ plane, as shown in Figure \ref{fig:regge}, consistent with the Regge theory.

\section{Conclusions}

In this work, we have shown that it is possible to extract meaningful information about the fundamental laws of nature directly from experimental data, without relying on prior theoretical knowledge, by leveraging the power of artificial intelligence. In particular, we demonstrated that even simple clustering algorithms can distinguish particles decaying via the strong interaction from those decaying via the weak interaction, solely by analyzing their lifetime distributions.
Linear dimensionality reduction techniques applied to decay modes allowed us to identify the baryon number as an emergent feature. Furthermore, by applying nonlinear dimensionality reduction to basic particle properties — spin, mass, and lifetime — we were able to reconstruct key properties of the flavor symmetries of the Standard Model, including isospin and SU(3) symmetry, as well as uncover indications of the existence of strange, charm, and bottom quantum numbers.
Additionally, our analysis of particle decay modes revealed structures consistent with baryon excitations organized along Regge trajectories, and we were able to recover the approximate universality of the Regge slope.

These results represent a promising step toward the broader goal of building aspects of fundamental physics from data using AI. Note that our approach has been essentially only kinematical. Much work needs to be done to use AI to uncover the dynamics of these systems, including the general structure of interactions in a Lagrangian formalism, amplitudes, etc.

Even before addressing challenging dynamical aspects, much remains to be done. A major future direction is to recover quarks as the basic building blocks of hadrons and to explore whether machine learning can infer more formal elements of quantum field theory, such as gauge symmetries.
Once we have verified that machine learning can reproduce known structures, the next natural question is whether it can go beyond them — identifying previously unrecognized patterns that might point toward new physics, such as hidden symmetries, evidence for new particles or the classification of the dozens of new hadrons discovered at LHC and other experiments in recent years (see for instance \cite{Ali:2017jda, Chen:2022asf}).%exotic states like the apparent tetraquarks and pentaquarks that may have been at CERN recently.

It is also worth noting that, in this study, the flavor symmetry clusters were associated with known multiplets but without identifying them as representations of a symmetry group. %using prior knowledge. 
A particularly interesting challenge is to train the model to recognize symmetry patterns and their mathematical representations autonomously, and to test whether it can correctly associate clusters with the appropriate symmetry groups — an approach in line with the previous works\footnote{We thank Sven Krippendorf for pointing out this possibility.} \cite{Krippendorf:2020gny,syvaeri2021improvingsimulationssymmetrycontrol,Wetzel:2020jan}. 
We leave the exploration of these and other open directions for future investigations.

\section*{Acknowledgements}

We thank Challenger Mishra, Francesco Muia and Andreas Schachner for general discussions and early collaboration. We are also thankful to  Steve Abel, Ben Allanach, Nicolás Bernal, Cliff Burgess, Michele Cicoli, Jim Halverson, Thomas Harvey, Luis Ib\'a\~nez, Vishnu Jejjala, Sven Krippendorf, Andre Lukas, Manuel Morales, Mario Ramos-Hamud, Christopher Thomas and Gonzalo Villa for interesting conversations. We thank Ben Allanach, Sven Krippendorf and Manuel Morales for comments and suggestions on a previous version of the manuscript. Our research is supported by a NYUAD research grant.

\bibliographystyle{utphys}
\bibliography{biblio}

\end{document}